\documentclass[aps,onecolumn,superscriptaddress]{revtex4-2}
\usepackage{amsmath,amsfonts,amsthm,amssymb,gensymb,float,graphicx,csquotes}
\usepackage{subcaption}
\usepackage{threeparttable}
\usepackage{hyperref}
\hypersetup{colorlinks=true,linkcolor=blue,urlcolor=red,citecolor=red}

\begin{document}

\title{Towards a Digital Twin for the Ground to QEYSSat Quantum Link}

\author{Henri P. N. Morin}
\affiliation{Institute for Quantum Computing, University of Waterloo, 200 University Ave W, Waterloo, Ontario, Canada, N2L 3G1}
\affiliation{Department of Physics and Astronomy, University of Waterloo, 200 University Ave W, Waterloo, Ontario, Canada, N2L 3G1}

\author{Alex Maierean}
\affiliation{Institute for Quantum Computing, University of Waterloo, 200 University Ave W, Waterloo, Ontario, Canada, N2L 3G1}
\affiliation{Department of Applied Mathematics, University of Waterloo, 200 University Ave W, Waterloo, Ontario, Canada, N2L 3G1}

\author{Brendon L. Higgins}
\affiliation{Institute for Quantum Computing, University of Waterloo, 200 University Ave W, Waterloo, Ontario, Canada, N2L 3G1}
\affiliation{Department of Physics and Astronomy, University of Waterloo, 200 University Ave W, Waterloo, Ontario, Canada, N2L 3G1}

\author{Ian DSouza}
\affiliation{Institute for Quantum Computing, University of Waterloo, 200 University Ave W, Waterloo, Ontario, Canada, N2L 3G1}
\affiliation{School of Physical and Chemical Sciences, University of Canterbury, Christchurch, New Zealand}

\author{Vinodh R. R. Muthu}
\affiliation{Institute for Quantum Computing, University of Waterloo, 200 University Ave W, Waterloo, Ontario, Canada, N2L 3G1}

\author{Thomas Jennewein}
\affiliation{Institute for Quantum Computing, University of Waterloo, 200 University Ave W, Waterloo, Ontario, Canada, N2L 3G1}
\affiliation{Department of Physics and Astronomy, University of Waterloo, 200 University Ave W, Waterloo, Ontario, Canada, N2L 3G1}
\affiliation{Department of Physics, Simon Fraser University, 8888 University Dr W, Burnaby, British Columbia, Canada, V5A 1S6}

\date{August 22, 2026}

\begin{abstract}
Simulations of physical systems require high-fidelity models to accurately represent reality. Simple models may be analytically tractable, but may not be sufficiently representative of reality for the given application. The cost of this simplicity is accuracy, or in the case of quantum key distribution, provable security. Sources of this accuracy gap include the difficulty of modelling physical effects which do not lend themselves well to analytical descriptions, such as afterpulsing. Here, we introduce a novel Monte Carlo based photon emission, transmission, and detection simulator, designed in the context of the Quantum Encryption and Science Satellite (QEYSSat) mission. Within this simulator, every major physical effect a photon may experience during an experiment, from emission to detection, can be accounted for in a probabilistic manner. This methodology allows for the inclusion of experimental parameters which are relevant for a satellite mission, and their impacts on secure key lengths. This simulator serves as a comprehensive baseline to predict and validate experimental data for the upcoming QEYSSat mission.

\end{abstract}

\maketitle

\section{Introduction}\label{sec:introdudction}

Quantum communications and quantum key distribution (QKD) simulations usually generate representative quantum bit error rate (QBER) values according to specified parameters, and is used to optimize the secure key length equations. However, they make simplifications, optimizing parameters which directly enter the algebraic secure key equations, rather than the detailed experimental parameters and behaviour \cite{NJP2013,strathclydesimulation,Khmelev2023}. Understanding the impact of the technical details and behaviour of the components and subsystems used in a satellite mission requires high-fidelity simulations, in order to predict expected observations to understand the detailed performance of the quantum link. 

The Canadian Space Agency's (CSA) Quantum Encryption and Science Satellite (QEYSSat) is the first Canadian satellite QKD mission \cite{CSAQEYSSat}. Notably different than other missions (e.g., \cite{micius,Li2025b,SpooQy,ZfT2024}), its main mission is set to perform uplink, rather than contemporary satellites predominantly performing downlinks, due to the advantages the latter configuration implies, e.g., turbulence mitigation \cite{Bonato2009,NJP2013,Lu2022}. Significant efforts have been produced to establish the feasibility of this mission, e.g., \cite{NJP2013,Bourgoin2015PRA,Bourgoin2015,Pugh2017,Anisimova2021}, as a detector in space presents systematically different challenges than a quantum source in space.

While low Earth orbit (LEO) satellite-based QKD offers a loss advantage compared to ground-based implementation, they suffer from limited connectivity. Ground-based implementations can continuously accumulate statistics e.g., \cite{Zahidy2024}, while satellite passes are severely limited, e.g., \textit{Micius} \cite{micius} being limited to five minutes. Therefore, predicting how much secure key can be distilled is of utmost importance. These predictions are constrained by both the experimental complexity, and the underlying security proof of the secure key equations. Simulating observations are an important part of these predictions. Generating as realistic observations as possible, will lead to more trustworthy and truly secure key distillation. Monte Carlo simulations have been applied to a wide range of applications in quantum physics \cite{Ceperley1986,Molmer1993}. The intrinsic randomness of the methods allows for the fine-grained simulation of the optical behaviour of the systems involved.

To this end we developed the QEYSSat Monte Carlo Simulator (QMCS), aimed at producing representative measurement data which mimics the detailed photon count data observed by the QEYSSat payload. Our simulator enables the detailed simulation of the main physical effects which are mostly stochastic, including the photon emission, quantum link, and photon detection, as expected for transmission of photons between ground and space. Our models in particular include detector afterpulsing and deadtime, which are memory effects and are difficult to implement analytically \cite{Ziarkash2018}. A visualization of the included quantum effects is given in Sec.~\ref{sec:methodology}. Directly simulating representative observations provides a variety of benefits. These include but are not limited to isolating the effects of low-level experimental parameters on the final key length, determining operational parameters for QEYSSat's detectors, analyzing the same experimental data with different secure key equations, and including different sources which underlie different physical mechanisms. Additional simulator features include the ability to simulate an uplink and a downlink, the ability to simulate a fixed link in addition to a variable link, and the automation of simulation and analysis.

To produce accurate observations through a free-space link, our software stack builds upon previous QEYSSat analyses e.g., \cite{NJP2013,Pugh2017,Yastremski2025}, to include an accurate model of the channel. These models include atmospheric effects obtained through, e.g., MODTRAN, and include diffraction, turbulence, absorption, and transmittance (details in Sec.~\ref{sec:linkmodel}). Representative orbital passes for QEYSSat are used in combination with these channel models to produce link models for the QMCS. Previous radiation studies have quantified the effect that radiation will have on the primary payload once in orbit, e.g., \cite{Anisimova2021}. Additionally, refs. \cite{Anisimova2017,DSouza2021} have shown that thermal annealing of the detectors will reduce the dark count rates at the cost of higher afterpulsing probability. Our detector models explicitly include the excess bias voltage of the detectors, allowing us to model expected behaviour of the detectors from the start to the end of the mission.

\begin{figure}[!htbp]
    \centering
    \includegraphics[width=0.45\linewidth]{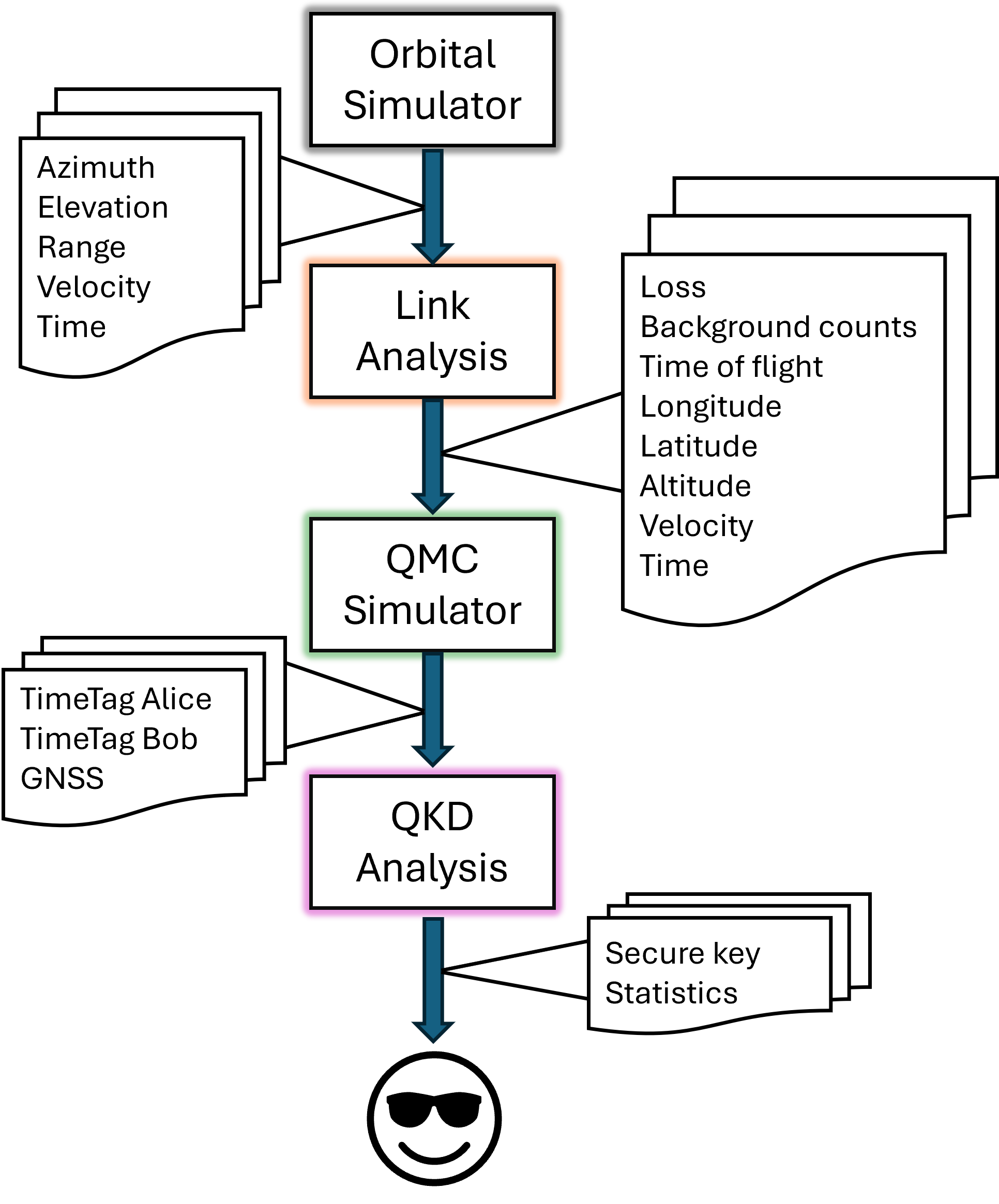}
    \caption{Full QEYSSat digital twin software stack. The data flow between each software is highlighted.}\label{fig:fullStack}
\end{figure}

The entire QEYSSat digital twin software stack is presented in Figure \ref{fig:fullStack}. It is a four-part model, beginning with orbital simulations, and terminating with the distillation of the secure key. The QEYSSat Monte Carlo Simulator itself works by generating a bit string representing the key that the sender wishes to emit, and then adding, subtracting, or otherwise altering the photon states according to functions modelling the various phenomena that affect the transmission process. Its purpose is to create representative files for the mission. This includes the proprietary data format and assuming experimental designs.

\section{Methodology}\label{sec:methodology}

\subsection{Introduction}
The simulator aims to replicate the experimentally generated timetag files from a quantum optics experiment. It is baselined to simulate a weak coherent pulse source (WCP) implementing decoy state BB84 using polarization encoding \cite{Schrier2026}, and an entangled photon source implementing BBM92 \cite{Islam2024}, which are QEYSSat's primary sources \cite{Jennewein2014}. QEYSSat's primary payload is its detectors \cite{Podmore2021}. These are silicon avalanche photodiodes (APD) operating in Geiger mode. These detectors will exhibit some afterpulsing, which will negatively contribute to the final secure key length. Furthermore, afterpulsing introduces memory effects which are hard to model analytically. For accurate simulated observations, these effects must be included. 

\begin{figure}[!htbp]
\centering
\includegraphics[width=\columnwidth]{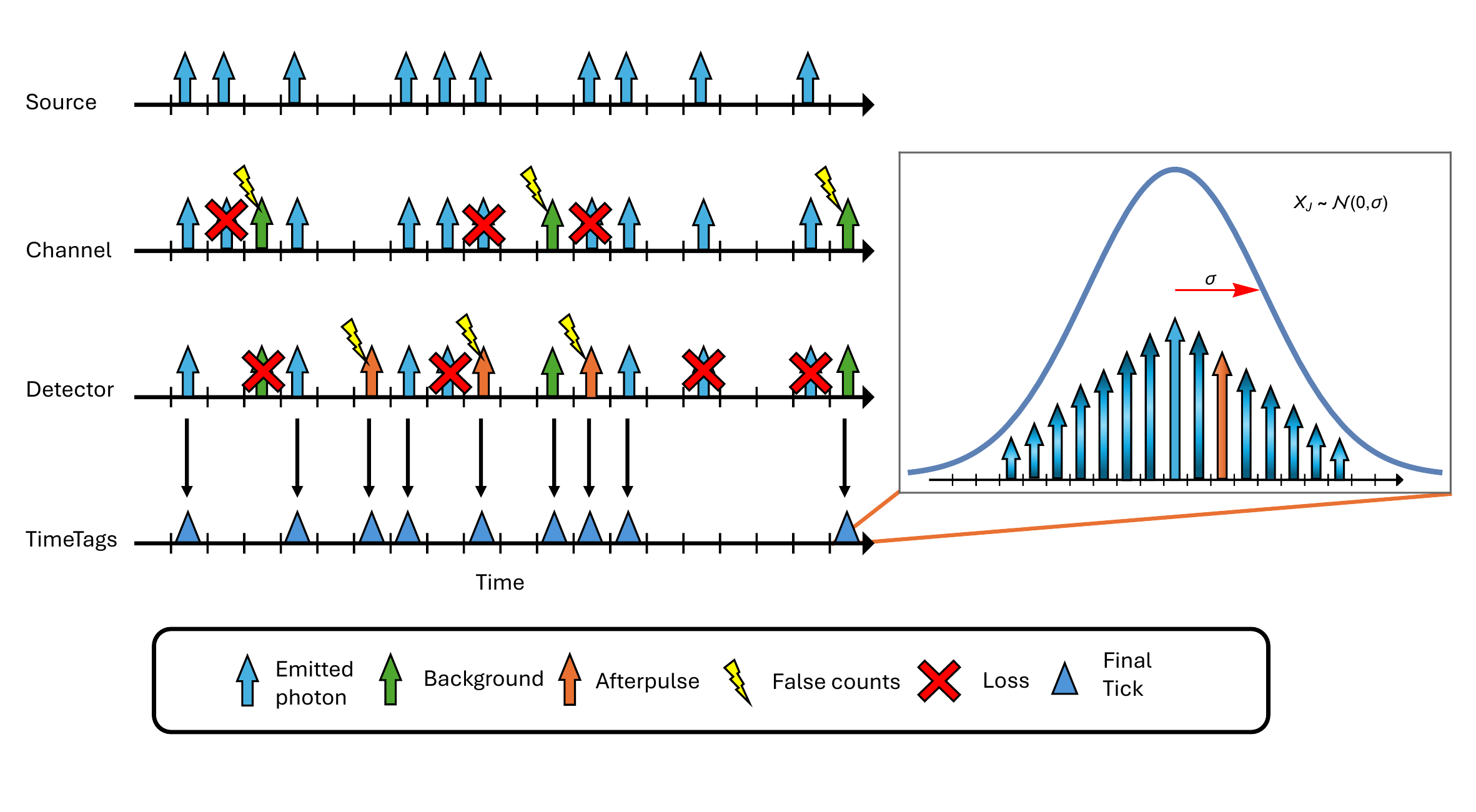}
\caption{Visualization of the effects modelled in our link simulation, and the different modules of the simulation chain. The simulation operates with a fine-grain time bin (78 ps), and each photon is associated with a certain timetag. The relevant modules add, subtract, or move timetags within the time bins. First, the source randomly generates emissions for each bin in the given timestep. Second, the channel module removes emissions due to loss, and adds emissions due to background counts. Third, the detector module removes emissions due to detector efficiency and deadtime, adds emissions due to dark counts and afterpulsing, and changes the time ordering of events due to timing jitter as a final step. Finally, once the jitter is applied, the identity of each tag is lost, and only a timetag remains. Each time step is simulated in full and written to files before starting over. The inset focuses on the timing jitter. The central emission is the true emission prior to applying the timing jitter. This will change the timetag to any of the other options within the distribution, scaled according to likelihood. In this example, the orange emission is the new timetag.}\label{fig:timeTagEffect}
\end{figure}

Therefore, the simulator is based on Monte Carlo methods, to probabilistically and temporally simulate every major physical effect which can affect an emitted photon, from emission to detection. This includes but is not limited to intrinsic source timing jitters, turbulence in the atmosphere, and detection efficiency, see Fig. \ref{fig:timeTagEffect}. We perform a finite timestep simulation, combined with randomized functions in order to mimic the quantum optical behaviour of the systems involved. The simulator is written in \texttt{MATLAB} and has been optimized to take advantage of the language's features.

\begin{figure}[!htbp]
    \centering
    \includegraphics[width=0.45\linewidth]{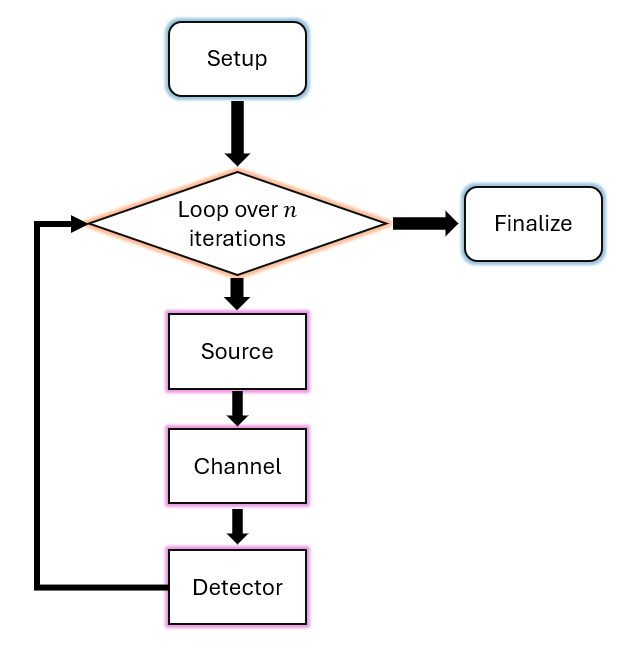}
    \caption{Main iteration performed by the QEYSSat Monte Carlo Simulator. The QMCS defines a given scenario to simulate based on given inputs. It iterates over the Monte Carlo time steps performing source tasks, channel tasks, and detector tasks, before safely saving all output files. The time steps are nominally 1 ms of real time.}
    \label{fig:simulateTimeStep}
\end{figure}

The simulator is structured as follows (see also Fig. \ref{fig:simulateTimeStep}). It first sets up the given scenario structures and objects, reading from externally defined files where relevant (e.g., link analysis files see Sec.~\ref{sec:linkmodel}). The simulator iterates over generating source photons, transmitting them through the channel, and detecting them at the receiver. The nominal timestep of this loop is $1$ ms. The system models two discretizing timetaggers, one at the source and the other at the receiver. These operate independently. The nominal timing resolution is 156.25 ps at the source, and 78.125 ps at the receiver, governed by internal clocks whose actual operational frequency deviates slightly over time, leading to clock drift - this effect is also incorporated.
 
The QMCS allows for a simple eavesdropper (Eve) to interact with the emitted photons from the source, within the free-space optical channel module. In principle, she can perform any attack allowed by quantum mechanics. To demonstrate a potential attack, our simple Eve implementation performs an intercept-resend attack, which is described in Sec.~\ref{sec:eavesdropper}. The unperturbed channel itself may still impact the emitted signals, by either polarization distortion, or in a free-space experiment, turbulence. As high loss is expected during a satellite pass, most photons are lost in the channel, which simplifies the computational complexity of the simulation.

The QMCS has been under development for many years \cite{DSouza2018,Maierean2024}. It is designed to be QEYSSat's digital twin, allowing full replication of a true satellite pass, prior to launch. This includes generating experimental data files in the same format. These are nominally analyzed through proprietary analysis software. Directly generating the timetags allows for additional analysis, beyond the coincidence and timing analysis required to distill a secure key.

QEYSSat's secondary payload includes a WCP source, to perform a downlink experiment, in contrast to the primary uplink. The simulator is able to simulate this configuration as well, highlighting the flexibility of our methods. Interested readers are directed to \cite{RefQSPIE} for additional details on the secondary mission.
 
\subsection{Sources}
To ascertain the security of the QKD protocol, determining the photon number distribution of the pulses emitted by a given source is critical. These photon numbers are generated in the source modules, and translated in timetags, which are tracked throughout the entire simulator. The \enquote{source} module performs the following tasks: 1) generate photon emissions 2) generate pulse per second synchronization pulses 3) apply Alice's intensity detector and 4) save data to files. 

Tracking the density matrix of each individual emitted state would be too computationally intensive. Instead, there are only a limited number of \enquote{states} which are tracked (as density matrices) in a state \enquote{palette}, and each emission is assigned an index into that palette. When certain processes such as depolarization is applied, it is applied to the states in the palette, affecting every emission that indexes that state. This simplification captures the relevant statistics, which are sufficient for our purposes.

The source modules generate photon numbers according to the source's probability distribution. However, it is important to note that the photon numbers themselves are not explicitly tracked outside of this module. Instead, $n$-photon states are treated as $n$ copies of the same timetag. As each individual timetag will be acted upon individually in subsequent operations, this combination will follow the expected statistics of a single $n$-photon state. Additionally, vacuum states are removed from tracking at the earliest possible moment, as there is nothing to detect and modify in further processing steps, additionally simplifying the computational task. 

To illustrate our methods and implementation, the weak coherent pulse source will be described in detail in the following section, for a single Monte Carlo timestep of 1 ms. 

\subsubsection{Weak Coherent Pulse}\label{sec:sourceWCP}
The weak coherent pulse source (WCP) is implemented as a heavily attenuated laser. It has a photon number probability distribution derived from treating the laser as a coherent state, namely the Poisson Distribution, which can be easily simulated:
\begin{equation}\label{eq:poisson}
    \text{Prob}(\alpha, n) = \frac{e^{-|\alpha|^2}|\alpha|^{2n}}{n!},
\end{equation}
where $|\alpha|^2$ is the mean photon number and $n$ is the photon number of a given pulse. This distribution has a continuous and infinite domain, which needs to be truncated for computational reasons. In addition, it is convenient to truncate it for practical reasons, since high photon numbers are usually not populated. To do so, we determine an appropriate length of photon numbers to consider for the current set of random numbers to generate, using the inverse cumulative distribution function of the Poisson distribution and the current mean photon number. The truncated probabilities are added into the last bin. For a signal intensity of $0.5$, the maximum photon number considered is $13$.

QEYSSat's WCP source runs at $400$ MHz and implements decoy state BB84 using polarization encoding \cite{Schrier2026}, enhancing the design from \cite{Pugh2017}. Thus, in a single timestep of 1 ms, 400 000 pulses are generated. Each of these emissions, for both the signal and decoy intensities, must be assigned a polarization. To implement secure QKD, the sequence of states used must be completely random. This requires a large ($>100$ GB), pre-generated set of random numbers for every satellite pass. As it is inconvenient to use during initial development and testing purposes, a repeating sequence mode was created. A random sequence of nominally $1000$ pulses is generated separately and ingested by the simulator while generating the WCP pulses. This sequence is therefore nominally repeated 400 times per timestep.

For a signal intensity of $|\alpha|^2 = 0.5$, the probability of emitting zero photons is $P(0)\approx 61\%$. For a decoy intensity of $|\alpha|^2 = 0.1$, $P(0)\approx 90\%$. In the nominal sequence file, $80\%$ of pulses are signal states, $14\%$ of pulses are decoys, and the remaining $6\%$ are vacuum. Overall, empty vacuum pulses make up 6\% of pulses, empty decoy pulses make up $12.7\%$ of pulses, and empty signal pulses make up 48.5\% of pulses. Removing all of these pulses (67.2\%) from the original $400\,000$ pulses leaves 32.8\% of pulses, or $\approx 131.2 \times 10^3$ pulses for the 1 ms chunk.

Of the non-empty pulses that remain, each has a random number of photons: single-photon pulses make up $77.8\%$, two-photon pulses $18.7\%$, three-photon pulses 3.1\%, and four-photon pulses 0.4\%, with higher photon numbers being ${\approx}\,0\%$. Calculating the probabilities for each photon number allows us to discuss the implementation while providing a quantitative way to validate against the actual photon numbers generated by the simulator.

After the random number of photons per pulse are generated, they are translated to the timetag and \enquote{state indices} variables. This tracks the timetag of emission, and which polarization its assigned to, respectively. Multiphoton events are represented by repeating the exact timestamp for as many photons which are in the given pulse. These timetags and state indices are the main variables tracked throughout each iteration.

As the WCP source implements the design from Ref. \cite{Pugh2017}, Alice locally monitors the pulses she is emitting. This is done with a beamsplitter, nominally siphoning off $0.1\%$ to her intensity detector. The simulator extracts these photons by sampling a uniform random distribution, and keeping the tag if the random number is below the given ratio. The pulses which are siphoned off are then detected through the same sequence of events as described in Sec.~\ref{sec:detectors}. In practice, the split off photons are used for polarization state monitoring.

Additionally, a marker \enquote{sync} pulse is emitted at each repetition of the non-random state sequence. For the 400 000 pulses in a 1 ms simulated pass, there are 400 marker pulses added to Alice’s output file. In total, Alice’s output data file should contain $\approx$ 500 signals for a single time step. A pair of these marker or \enquote{sync} pulses indicate that a \enquote{segment} of laser pulses were emitted. The number of pulses between two sync pulses is known. This information is used during analysis to estimate when emissions took place, which previous studies have found to be quite accurate, e.g., \cite{Pugh2017}, and it requires considerably less data collection and storage than recording every emission. These marker pulses are used during analysis in proprietary software to line up each timetag sequence. This additional software, the QKD Analysis Software in Fig. \ref{fig:fullStack}, takes its roots in \cite{MeyerScott2011}, with additional algorithms described in \cite{Bourgoin2015PRA,Higgins2020}.

\subsection{Channel}\label{sec:channel}
Once the photons are emitted by the source, they are transmitted to the channel module. Several effects are applied. If Eve exists, this is where she has an impact (see Sec.~\ref{sec:eavesdropper}). 

The \enquote{channel} module performs the following tasks: 1) conditionally apply Eve 2) apply loss 3) update state palette given polarization and rotation effects and 4) add background counts. Link loss and background counts are modelled separately for the given pass, and ingested by the simulator, as noted in Fig. \ref{fig:fullStack}. Description of this data is presented in Sec.~\ref{sec:linkmodel}. This data is generated at per second intervals and is therefore interpolated to match the designated timestep. This feature allows us to consider different ground stations and different link budgets simultaneously, helping make the operational decision as to which ground station gets access on a given pass.

The satellite pass use case expects $>35$ dB of loss, which significantly reduces the amount of information tracked. The random amount of background counts are sampled from a Poissonian distribution given by the ingested background rate and the time step. The generated counts are then randomly assigned a timetag. The background is unpolarized and a new state is added to the palette. Depolarization effects are applied on the state palette density matrices directly through their $\chi$ matrix representation. It is parameterized with its own parameter, which can be tuned during simulations. Additional effects can be added by adding their $\chi$ matrix representation.

Once the channel effects are applied, the photons are transmitted to Bob's detector module. Evidently, the satellite requires the implementation of a link whose loss varies over time. During development, a simpler fixed link was also developed. 

\subsubsection{Variable and Fixed Links}\label{sec:variableLink}
In the nominal uplink scenario, Alice is on the ground, sending her signals to Bob (the satellite). Alice records her Global Navigation Satellite System (GNSS) data and saves it to a separate dedicated file on a per-second basis. This information is used in the timing analysis in the proprietary QKD Analysis Software. 

Bob's GNSS data is originally generated in Ansys Systems Toolkit (STK), and processed before being ingested by the simulator, as noted in Fig. \ref{fig:fullStack}. It contains position (longitude, latitude, altitude) and velocity information on a per-second interval, matching what we expect to get from the satellite in orbit. As with the link data, the ingested orbital data is then interpolated to match both the desired simulation length and the given timestep resolution. At every timestep, the simulator takes the next element in these interpolated lists.

The variable link described above models the expected path of the satellite and is required to twin the mission. However, the moving receiver may hide bugs during development. Thus, a fixed link mode was simultaneously developed, where Bob does not move and the channel loss does not change over time. To use the same functions as the variable link, the fixed link mode generates its own link file, according to the same format. The variables within are now constant with respect to time. This feature is useful to simulate a fixed link on the ground.

\subsubsection{Eavesdropper}\label{sec:eavesdropper}
While a practical adversary would not be restricted in their attack vector, to demonstrate the impact of Eve, the simulator can add an eavesdropper into the channel module. She directly interacts with the emitted signals in the channel. The user can tune where Eve is placed in the channel (as a fraction of the distance between Alice and Bob), how big of a fraction of the intercepted signals she can measure, and how much delay her measurement is adding to the photon's time of flight. It is assumed the photons she splits off are used to perform a perfect measurement, where she gets full information when she chooses the correct basis. 

In an idealized attack where Eve intercepts all of the emitted photons which reach her, we can expect to observe a 25\% QBER. As this intercepted fraction can also be tuned, we can quantify Eve's presence on the overall QBER.

\subsection{Detectors}\label{sec:detectors}
QEYSSat's primary payload is a 4-state analyzer \cite{Podmore2021}. We model the analyzer from the initial passive beam splitter. Each detector is modelled identically, but can be parameterized separately. The \enquote{detector} module performs the following tasks: 1) apply detection efficiency, 2) add dark counts, 3) add afterpulses from the previous iteration if they exist, 4) applies intrinsic detector effects (afterpulsing, recharge time, deadtime, jitter) for current iteration 5) generates Bob's pulse per second synchronization pulses 6) generate and embed GNSS data within the timetags, and 7) save data to file. Each task is described in the following subsections.

\subsubsection{Detector Sorting, Efficiency, and Dark Counts}\label{sec:sortEffDCR}

The single timetag stream is now split into a stream for each detector. The passive basis choice non-polarizing beam splitter and the polarizing beam splitters are assumed to be lossless. They can nonetheless be made imperfect. The state assignment is made randomly, using a probability look up table from eligible detectors for a given timetag. 

For each detector the efficiency is applied. A uniformly distributed random number is generated, and if the number is less than the efficiency, the detector has been deemed to click and the tag is kept. Recharge effects are applied in a subsequent step (see Sec.~\ref{sec:RechargeDeadtimeJitter}). In a similar fashion as the initial emissions, the dark counts are added. First, the number of dark counts to add in the given timestep is determined by randomly sampling a Poissonian distribution whose parameter is given by the product of the dark rate of the detector and the given timestep duration. Then, this number of dark counts are then randomly placed (uniformly) within the given timestep. 

\subsubsection{Afterpulsing}\label{sec:afterpulsing}
After dark counts are added, afterpulses from a \textit{previous} iteration are added to the tracked list of timetags, prior to computing the \textit{next} round of afterpulsing. The inclusion of afterpulsing in our detector model is a novel part of our simulator. An afterpulse is when one avalanche in a detector generates a new one, even if no new pulse arrived to the detector. These therefore depend on the prior state of the detector. With a high enough afterpulsing probability, long chains are possible. This induces correlations between detections which must be constrained for the given QKD security proof to apply. Additionally, they are additional extraneous counts which increase noise and worsen detection statistics. 

Whether a given detection event triggers an afterpulse is parameterized by the afterpulsing probability $p_{ap}$. This can be viewed as the probability that exactly one electron is trapped within the detector substrate in a given avalanche, under the assumption that at most one electron gets trapped during any given avalanche. Such an electron will eventually get detrapped. The detrapping time is random in nature, and its probability distribution is modelled by
\begin{equation}\label{eq:afterpulsing}
    P_{\text{de-trap}}(t)=\lambda_{ap} e^{-\lambda_{ap} t},
\end{equation}
where $\lambda_{ap}$ is the characteristic rate of afterpulsing and $t$ is the time elapse since the electron got trapped. As each detection event \textit{could} trigger an afterpulse, these must be computed once all new counts are added \cite{DSouza2018}. Sampling from this distribution is performed through inverse transform sampling. 

Furthermore, each afterpulse itself can generate a new afterpulse, according to the afterpulsing probability $p_{ap}$. Our algorithm allows generation of unrestricted afterpulsing chains \cite{DSouza2018}. Sometimes, these chains exceed the given simulated timestep. These must be added to the tracked list of timetags during the subsequent iteration of the main loop. 

\subsubsection{Recharge Time, Deadtime, and Timing Jitter}\label{sec:RechargeDeadtimeJitter}
Once all possible detection events are \textit{added}, we need to start \textit{removing} certain events. This is due to both the recharge time of the detector, and the deadtime after a detection. As noted above, afterpulsing times are explicitly carried between iterations of the main loop, as the unrestricted chains are not guaranteed to be fully contained within a single timestep. Thus, if an avalanche which caused an afterpulse is removed by the recharge time or the deadtime, so are all linked afterpulses. 

As noted in Sec.~\ref{sec:sortEffDCR}, we initially kept every detection event which survived the efficiency calculation. However, certain pulses might have been created during the detector deadtime and thus are removed at this stage. The physical detectors are passively quenched, and thus have a longer recharge time to peak efficiency than actively quenched detectors. The deadtime is modelled as: 
\begin{equation}\label{eq:defDeadTime}
    t_{dt} = - \tau * \ln(1 - \frac{V_{Th}}{V_{PH}}),
\end{equation}
where $t_{dt}$ is the deadtime, $\tau$ is the recharge time, $V_{Th}$ is the discriminator threshold and $V_{PH}$ is the pulse height of the detector. A detection only happens when the output voltage is above the threshold. The recharge time $\tau$ quantifies the time it takes for the detectors to return to ready state. In the physical detectors, the output voltage after a click increases according to:
\begin{equation}\label{eq:defRechargeTime}
    V(t) = V_0 (1-e^{\frac{-t}{\tau}}),
\end{equation}
where $V_0$ is the normal output voltage for an avalanche associated with the full detector efficiency, and $t$ is time after a click. This affects the detector efficiency which follows a double exponential \cite{DSouza2018}:
\begin{equation}\label{eq:rechargeDoubleExp}
    \eta(t) = \dfrac{1}{1-e^{-\frac{V_0}{V_c}}} \big(1-e^{-\frac{V(t)}{V_c}}\big)
\end{equation}
where $V_c$ is a characteristic voltage for the given diodes. This expression asymptotes to 100\% efficiency, as the detector efficiency itself has already been accounted for.

Until this step, the list of detection events are exactly ordered in time and assumed to take place at an exact time. However, that is not representative of true detectors. As a final step modifying the timetags, timing jitter is applied. Timing jitter is modelled by a normal distribution with a given standard deviation $\sigma$. It is implemented by generating a sufficient number of random numbers by sampling a standard normal distribution, which are then multiplied by the parameterized timing jitter (see Fig. \ref{fig:timeTagEffect}).

The code then generates pulse per second (PPS) and GNSS data for Bob and embeds them into his timetag file, before saving the data and going to the next iteration of the loop, as indicated in Figure~\ref{fig:simulateTimeStep}.

\subsection{Link Modelling}\label{sec:linkmodel}

\begin{figure}[!htbp]
\centering
\includegraphics[width=0.90\columnwidth]{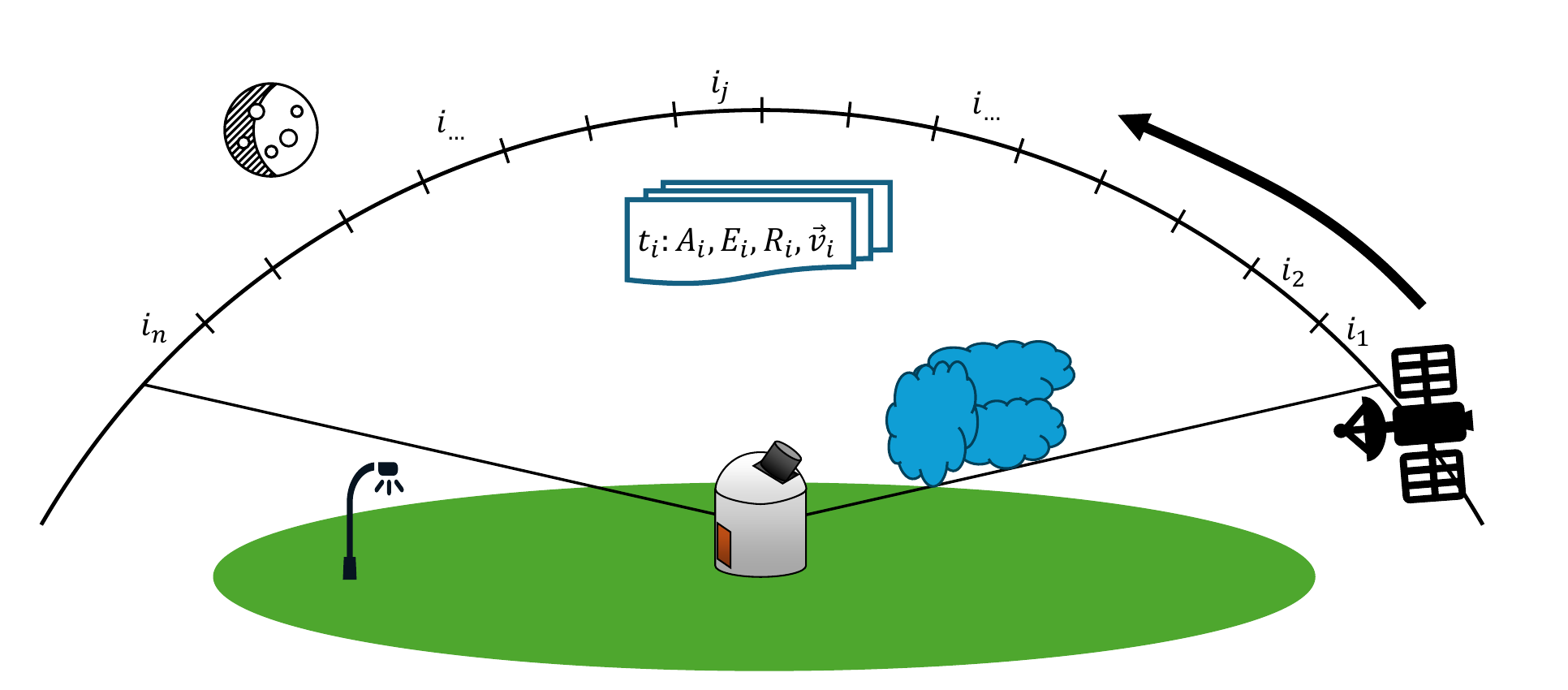}
\caption{Discretized orbital pass. At each second $t_i$ of the simulated pass, azimuth $A_i$, elevation $E_i$, range $R_i$, and velocity $\vec{v}_i$ data are generated within Ansys STK. This information is used to compute link properties as a function of time which is in turn ingested by the QEYSSat Monte Carlo Simulator. This data is assumed to be the quantum phase of the total orbital pass. Considered effects include turbulence, reflected moonlight, and light pollution.}\label{fig:linkModelPerTime}
\end{figure}

\begin{table}[!htbp]
\begin{center}
\begin{tabular}{|c|c|}
\hline
Parameter & Value \\
\hline
Altitude & 550 km \\
Acceptance cone & 45$^\circ$ \\
Orbit type & Polar, sun-synchronous \\
Receiver aperture & 25 cm \\
Transmitter aperture & 50 cm\\
Wavelength & 785 nm\\
$C_n^{(2)}(0)$ & $1.7*10^{-14}$ m$^{-2/3}$ \\
Wind & 21 ms$^{-1}$ \\
Internal optic loss & 3 dB\\
Pointing error & $2\,\mu$rad \\
\hline
\end{tabular}
\caption{Relevant orbital and link parameters used to generate the example pass used in the results presented in Figure~\ref{fig:qberSinglesHist}.}\label{tab:baselinePassOrbitalParams}
\end{center}
\end{table}

The QMCS ingests data generated from the Link Analysis (as shown in Fig. \ref{fig:fullStack}). Specifically, the simulator requires link loss (dB), background counts (cps), photon time of flight (s), and the satellite's latitude (degree), longitude (degree), altitude (km), and velocity (ms$^{-1}$) in the North-East-Down reference frame, all generated at per second intervals. Table \ref{tab:baselinePassOrbitalParams} highlights relevant characteristics for both the orbital simulation and the subsequent link analysis.

The modelling starts in Ansys STK with orbital simulation of a number of passes over a selected ground station. Azimuth, elevation, range, access, and velocity information are exported, at per second increments, following the data flow highlighted in Fig.~\ref{fig:fullStack}. This data is processed and passed through the Link Analysis software. This software expands on previous work \cite{NJP2013,Mohammadi2021} to calculate link loss, background counts, and photon time of flight. Additionally, it splices the position and velocity information to the selected pass, to generate two files for the simulator. One contains the link information, the other contains orbital information. Figure~\ref{fig:linkModelPerTime} highlights the concept for the link analysis itself. Rather than performing an analysis over a given elevation and translating the secure key equations, or other parameters, as a function of this elevation, the remaining characteristics are calculated at per second intervals using this orbital data. 

The photon time of flight is calculated from the satellite's and ground station location defined with the WGS84 ellipsoid. The geometric distance is then corrected for refraction induced delay \cite{Marini1973}. The temperature, pressure, and atmospheric humidity data are taken from measured data from the ground station at the University of Waterloo.

The link loss is calculated using the method described in \cite{NJP2013}. The background counts are currently calculated using the method described in \cite{NJP2013}. Future work will update this to use the method in \cite{Yastremski2025}.

Our modular data pipeline allows these methods to be updated independently from the simulator itself. Different analyses can be performed and simulated upon to predict their impacts on the secure key length. We expect to use this functionality to compare experimental data from QEYSSat.

A satellite pass is comprised of different phases, as tracking must be successful before any quantum information is transmitted. The simulator assumes the ingested pass information comprises the entirety of the quantum phase. Thus the simulator itself is agnostic to the given start and end points of the quantum phase of the pass. For the main example pass analyzed in Sec.~\ref{sec:results}, the quantum phase occurs in a 45-degree elevation cone. Additional post-processing may reject detection events based on e.g., QBER, from the higher loss elements at the beginning and end of the pass, essentially restricting the quantum phase \cite{Erven2012} after the fact. 

\section{Theoretical Description}\label{sec:theoryMC}

The QEYSSat Monte Carlo simulator is an innovation among existing QKD models, all of which tend to rely on \textit{analytic} descriptions of the transmission process \cite{strathclydesimulation, openQKDSecurity, Wang2019}. The simulator, on the other hand, passes each pulse from a bit sequence individually -- in the style of a Monte Carlo model, accounting for statistical effects such as detector inefficiency or depolarization from the atmosphere. 

Here we introduce \textbf{stochastic processes}, $\{X_n,\; n \in \mathbb{N}\}$, to describe the QMCS. In particular, a specific sub-type of random process called the \textbf{discrete-time Markov Chain} which is characterized by the \textit{memoryless property}, often also called the Markov property: the transition to the next state depends only on the current state and is not conditional on any other prior state. This is defined using the conditional transition probability as,
\begin{equation}
\begin{split}
P_{ij} &= P\{X_{n+1}=j \; | \; X_n=i, X_{n-1}=i_{n-1}, \dots, X_1=i_1, X_0=i_0\} \\ &= P\{X_{n+1}=j \; | \; X_n=i\}.
\end{split}
\end{equation}
In the QMC simulator, the random variable whose state is transitioning over time is the emission generated by the sender. The state of each pulse is a vector $(p, t, N)$, where the dimensions represent \textbf{polarization}, a vector on the Poincar\'{e} Sphere, \textbf{timestamp at emission} and \textbf{number of photons} in the pulse, both whole numbers. As the pulse propagates through the system and encounters a sequence of physical phenomena, at each point it may undergo one of four possible transitions:
\begin{enumerate} 
    \item The pulse is lost (undergoes ``annihilation"). 
    \item Effect on \underline{one} coordinate of the state vector (a ``transition").
    \item No change (a ``null" transition).
    \item A branching event occurs, producing an additional pulse, e.g., due to afterpulsing.
\end{enumerate}
The state of each pulse generated by the source, $\overset{\rightharpoonup}{X}_n = (p_n, t_n, N_n)$, evolves according to probability distributions associated with the underlying physical processes, which in aggregate can be thought of as a function, $h(\overset{\rightharpoonup}{X}_n)$. Due to the random nature of many of the effects in the simulator that go into defining $h$, it is not possible to write an analytic expression for it. In such situations where it is difficult to evaluate a model $h(\overset{\rightharpoonup}X)$, we turn to the \textbf{Markov Chain Monte Carlo (MCMC)} method and instead approximate: 
\begin{equation}
    \theta = \mathbb{E}[h(\overset{\rightharpoonup}X)] = \sum_{j=1}^\infty h(x_j)P\{\overset{\rightharpoonup}X=x_j\}.
\end{equation}
Repeated simulation of large ensembles of pulses then produces statistical distributions of experimentally relevant quantities such as QBER and final secure key lengths.

A level of sophistication and, consequently complication is presented in the simulator's stochastic model of physical effects where it characterizes the single-photon avalanche diode detectors in unprecedented detail \cite{DSouza2018}. Among the list of detector parameters is \textit{afterpulsing}: the occurrence of latent detections as a result of released charges that were trapped in the junction depletion layer of the detector during a prior avalanche. The timing of when charges get de-trapped follows a probability distribution, meaning that some afterpulses will happen immediately after the detection of the original incident photon (during deadtime), others after the deadtime (triggering another detection), and others yet may be released after many detections have been recorded since the original avalanche. Moreover, afterpulses can themselves cause further afterpulses, so-called higher-order afterpulses. Therefore, afterpulsing constitutes a \textit{memory effect} in the context of Markov chains. This challenges both the \underline{Markovian property}, as well as the \underline{Monte Carlo} nature of the model. In-fact, afterpulsing is labelled ``non-Markovian", meaning that it does not have first-order Markov memory. This is addressed neatly in \cite{Wang2016} where the authors propose an analytical expression for the afterpulsing probability including higher order afterpulsing. Additionally, afterpulsing makes for non-independent and identically distributed random variables (pulses) due to this memory effect, which challenges the use of the Strong Law of Large numbers, needed to define parameter estimation in the Monte Carlo setting. This, however, is addressed in \cite{ross2014} using an ergodic theorem.

\section{Results}\label{sec:results}

The timetag files produced by the QMCS are passed through proprietary analysis software, whose detailed description lies outside the scope of this work. Certain algorithms have been introduced in \cite{Bourgoin2015PRA}. It performs the timing and coincidence analysis, and generates additional files with the desired results, e.g., secure key file, QBER over time, and other useful statistics. Initial results presented are based on a baseline, optimal pass with the WCP source. 

\begin{figure}[!htbp]
\centering
\includegraphics[width=\columnwidth]{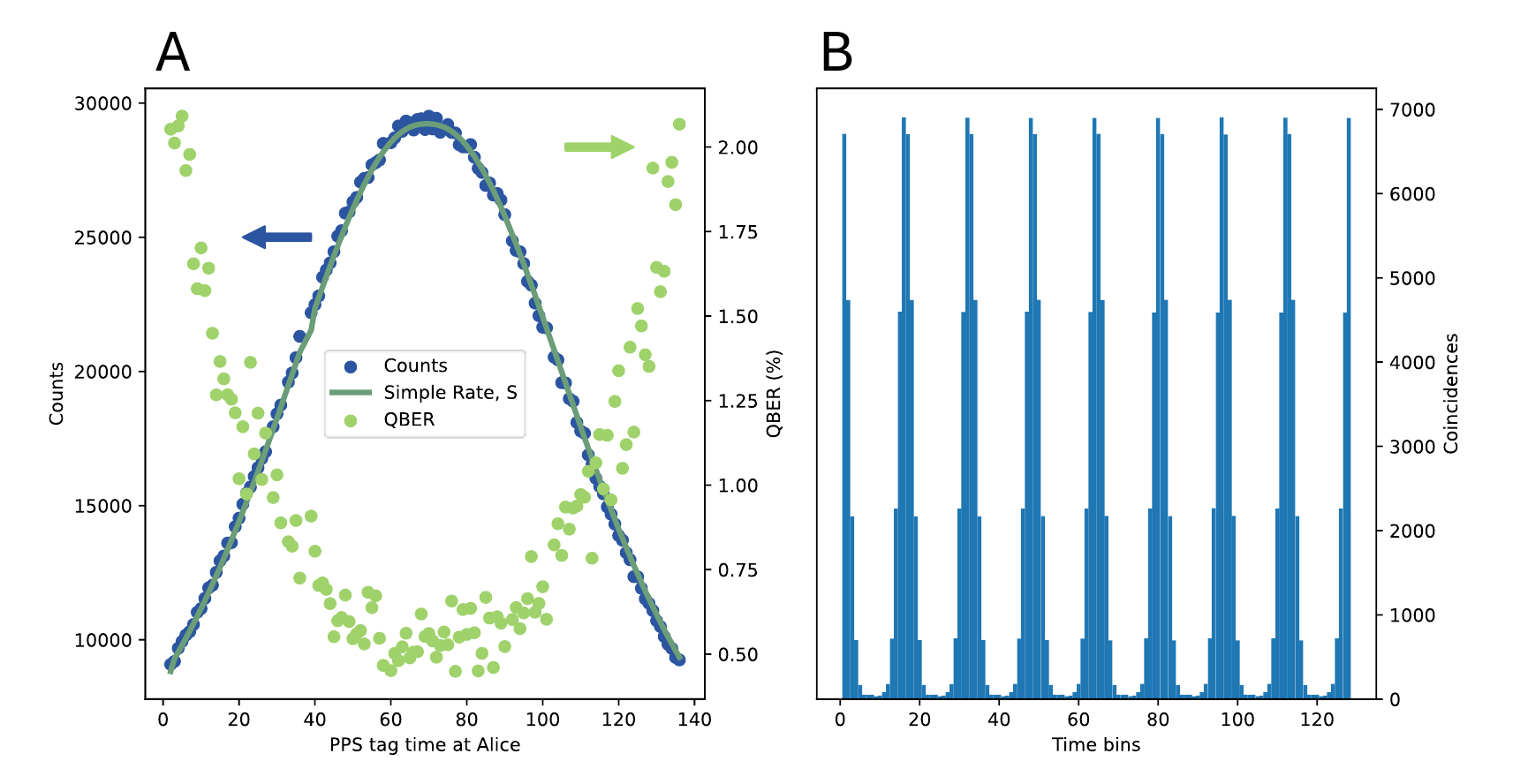}
\caption{QBER, singles counts, coincidence histogram for an example simulated pass of 140 s. (A) Single counts (left axis, blue) show an increasing amount of detected photons as the satellite approaches zenith. The highest concentration of counts is located at zenith, and correspondingly relate to the lowest QBER (right axis, green). QBER worsens dramatically at the edges of the pass, corresponding to the lowest elevation and longest transmission distance. In the middle of the pass, QBER is at its lowest, around 0.5\%. The simulated counts are compared against a simple model given by eq.~\eqref{eq:simpleRate}. (B) Coincidence histogram extracted from frame 63 of Figure~(A), which has the highest number of singles detected. Twenty (20) time bins are plotted, with the resolution of the timetagger (78.125 ps). The WCP source's Gaussian pulse shape and repetition rate can be visually identified. }\label{fig:qberSinglesHist}
\end{figure}

The simulated example satellite pass used in subsequent analysis is a close to zenith pass in December 2026 over QEYSSat's primary ground station at the Canadian Space Agency (CSA) in St.-Hubert, QC. Measured QBER, singles count, and coincidences are presented in Fig.~\ref{fig:qberSinglesHist}. Detector parameters used in this simulation are presented in Table~\ref{tab:nominalDetParam} and nominal source parameters in Table~\ref{tab:nominalSourceParam}. Each detector can be individually addressed, rather than approximating all detectors to be identical. Additional information is presented in Table~\ref{tab:passSpecs}.

\begin{table}[!htbp]
\centering
\begin{threeparttable}
\caption{Representative detector parameters used for highlighted results in Figure~\ref{fig:qberSinglesHist}. Other than the deadtime which is computed according to eq.~\eqref{eq:defDeadTime}, all other parameters are set by the end user.}\label{tab:nominalDetParam}
\begin{tabular}{|c|c|c|c|c|}
\hline
Parameter &  H Detector & V Detector & D Detector & A Detector \\
\hline
Efficiency$^{|}$ (\%) & 70 & 70 & 70 & 70\\
Timing jitter (FWHM)$^{\perp}$ (ns) & 0.4 & 0.4 & 0.4 & 0.4\\
Recharge time$^{\ddagger}$ (ns) & 720 & 700 & 690 & 710 \\
Pulse height$^{*}$ (V) & 1.1 & 1.1 & 1.2 & 1.1 \\
Discriminator threshold$^{\dagger}$ (V) & 0.05 & 0.05 & 0.05 & 0.05 \\
Deadtime$^{\S}$ (ns) & 34 & 34 & 30 & 32\\
Afterpulsing probability$^{\dagger}$ (\%)  & 0.8 & 3 & 4 & 1 \\
Afterpulsing characteristic rate$^{\ddagger}$ (s$^{-1}$)  & $10^6$ & $10^6$ & $10^6$ & $10^6$\\
Dark rate$^{||}$ (Hz) & 25 & 50 & 50 & 50 \\
\hline
\end{tabular}
\begin{tablenotes}
\item {$|$ Excelitas SPCM-NIR Datasheet \cite{ExcelitasNIR}}
\item {$\perp$ \cite{Podmore2025a}}
\item {$\ddagger$ \cite{DSouza2018}}
\item {* Private correspondence}
\item {$\dagger$ \cite{Anisimova2021}}
\item {\S~Equation \eqref{eq:defDeadTime} using other values from this table}
\item {$||$ \cite{DSouza2021}, assuming early stage of the mission with low radiation damage.}
\end{tablenotes}
\end{threeparttable}
\end{table}

\begin{table}[!htbp]
\begin{center}
\begin{tabular}{|c|c|}
\hline
Parameter & Value \\
\hline
Repetition Rate & 400 MHz \\
Signal mean photon number & 0.5 \\
Decoy mean photon number & 0.1 \\
Vacuum mean photon number & 0 \\
Laser jitter & 169 ps \\
Marker pulse jitter & 62 ps \\
\hline
\end{tabular}
\caption{Nominal WCP source parameters used for Figure~\ref{fig:qberSinglesHist} \cite{Schrier2026}.}\label{tab:nominalSourceParam}
\end{center}
\end{table}

\begin{table}[!htbp]
\begin{center}
\begin{tabular}{|c|c|}
\hline
Parameter & Value \\
\hline
Alice TimeTag File Size & 138 MB \\
Bob TimeTag File Size & 9.05 MB \\
Total number of detections & 2 583 440 \\
Pass length & 140 s \\
Simulation run time & 43 min - 80 min \\
Average QBER & 0.964 \% \\
Coincidence analysis and sifting & 43 s \\
Full QKD post-processing (EC + PA) & 350 + 20 s \\
\hline 
\end{tabular}
\caption{Analysis specs for example pass of Figure~\ref{fig:qberSinglesHist}. Simulation is performed on a compute cluster and analysis is performed on a local machine. The TimeTag files are in a proprietary compressed binary format to save storage space. After processing, Alice's TimeTag file is 1.01 GB and Bob's is 43.5 MB.}\label{tab:passSpecs}
\end{center}
\end{table}

The simulated QBER is below 2\% over the entire duration of the pass Fig. \ref{fig:qberSinglesHist}~(A). A maximum of approximately 30 000 single counts can be measured at zenith. This region has the lowest measured QBER of 0.5\%, making it the most useful section of the pass to generate the highest amount of secure key. The mean QBER over the entire pass is 0.964\%. QBER is calculated using all detected coincidences, for illustrative purposes. 

Using the approach and receding parts of the pass will be an optimization problem for the secure key analysis. Variable length security proofs will be able to tolerate this changing loss to maximize the distilled key \cite{Tupkary2026}. The single photon count rate obtained generated from the QMCS is compared against a simple link model given by:
\begin{equation}\label{eq:simpleRate}
    S = (R * \mu_{\text{total}} * \eta_{\text{loss}} * \eta_{\text{detector}} + \text{BG} + \text{DCR}) \,\Delta t,
\end{equation}
where $S$ is the expected singles count, $R$ is the repetition rate of the source (400 MHz), $ \mu_{\text{total}}$ is the total emitted mean photon number (0.414), $\eta_{\text{loss}}(t)$ is the efficiency of the channel (as determined by the link analysis, 36–41 dB), $\eta_{\text{detector}}$ is the detector efficiency (0.7), $\text{BG}(t)$ is the background counts reaching the detector (as determined by the link analysis, 110-800 cps), DCR is the intrinsic dark count rate for the detector (40 Hz), and $\Delta t$ is the integration time (1 s). This simple model necessitates assuming a single value for the detector parameters, which is less realistic. Nonetheless, Figure \ref{fig:qberSinglesHist}~(A) shows very good agreement between the simple model and the Monte Carlo simulation. 

Figure \ref{fig:qberSinglesHist}~(B) presents the coincidence histogram for the same baseline pass, specifically taking a one-second frame, 63 s out of 140 s. The bin size is the time tagger resolution of $156.25$ ps. The coincidence window is based on a loss heuristic from \cite{MeyerScott2011,Bourgoin2015PRA}. For the given simulation, it changes from 1.1 to 1.2 ns based on the estimated channel loss for a given frame of data, as determined from that reference.

\begin{figure}
    \centering
    \includegraphics[width=\linewidth]{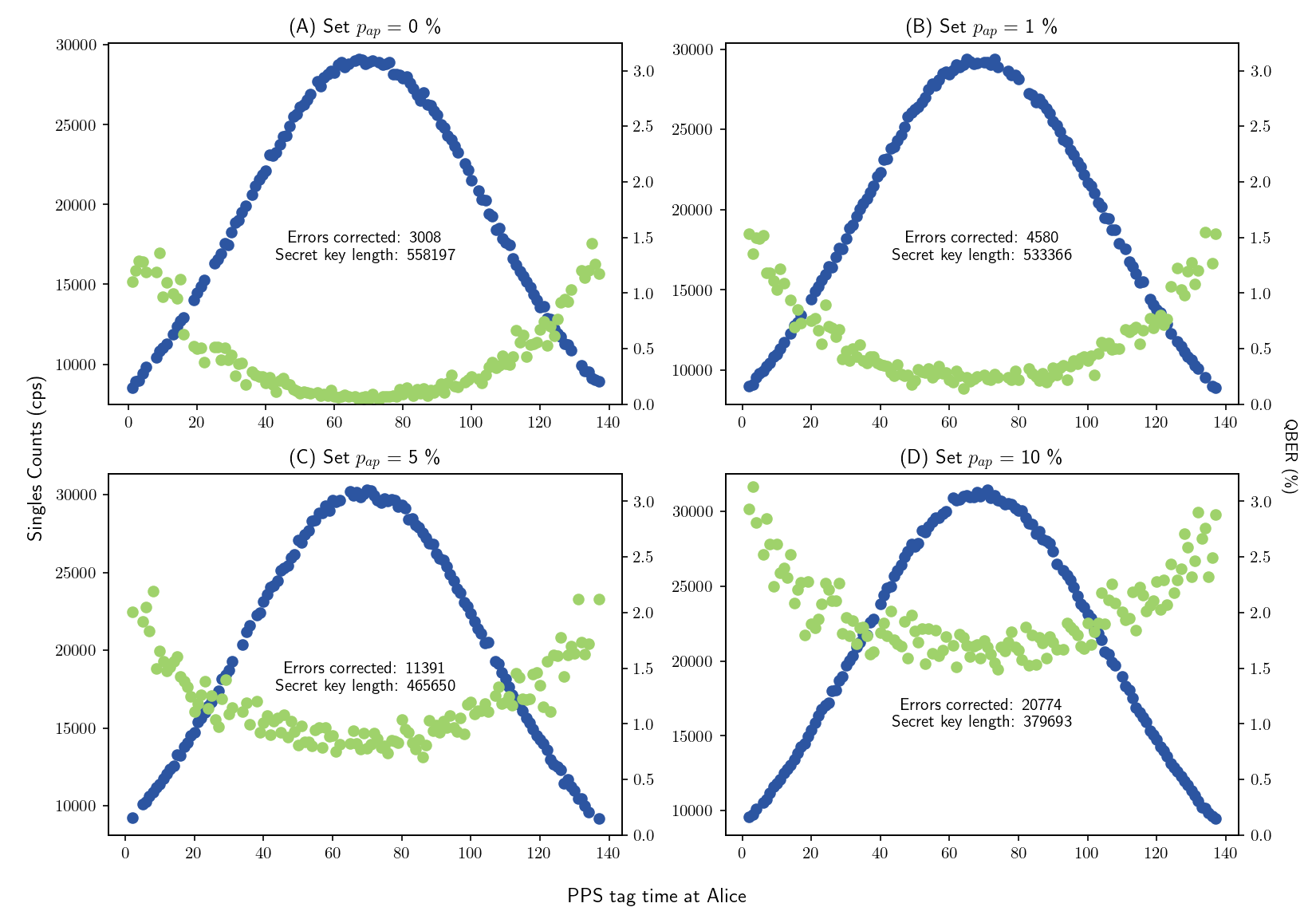}
    \caption{Afterpulsing effects on the secure key length. Different levels of afterpulsing are simulated for otherwise identical scenarios. As the afterpulsing probability increases, the QBER gets worse and has a wider spread of values across the pass. Moreover, the number of errors and the distilled secure key length get worse with increasing afterpulsing probability. In each subfigure, the single counts are in blue (left~axis) and the QBER is in green (right~axis).}
    \label{fig:afterpulsingKeyLength}
\end{figure}

Figure~\ref{fig:afterpulsingKeyLength} shows the effect of increasing afterpulsing probability on relevant QKD metrics, QBER and the distilled secure key length. To isolate the effect of afterpulsing, all detector parameters were made uniform across all detectors, unlike the parameters listed in Table~\ref{tab:nominalDetParam}. This includes the afterpulsing probability. Figure~\ref{fig:afterpulsingKeyLength}~(A) shows the ideal scenario, with $p_{ap}=0\%$. This analysis distills the highest key with the lowest amount of errors, which is expected. As the afterpulsing probability increases, the performance of the link deteriorates, with the QBER and number of errors getting worse. Future afterpulsing analysis will determine the observed afterpulsing probability based on simulated dark counts, without signal counts e.g., \cite{DSouza2018,Anisimova2021}.

\begin{figure}[!htbp]
\begin{subfigure}[b]{0.49\textwidth}
\includegraphics[width=\textwidth]{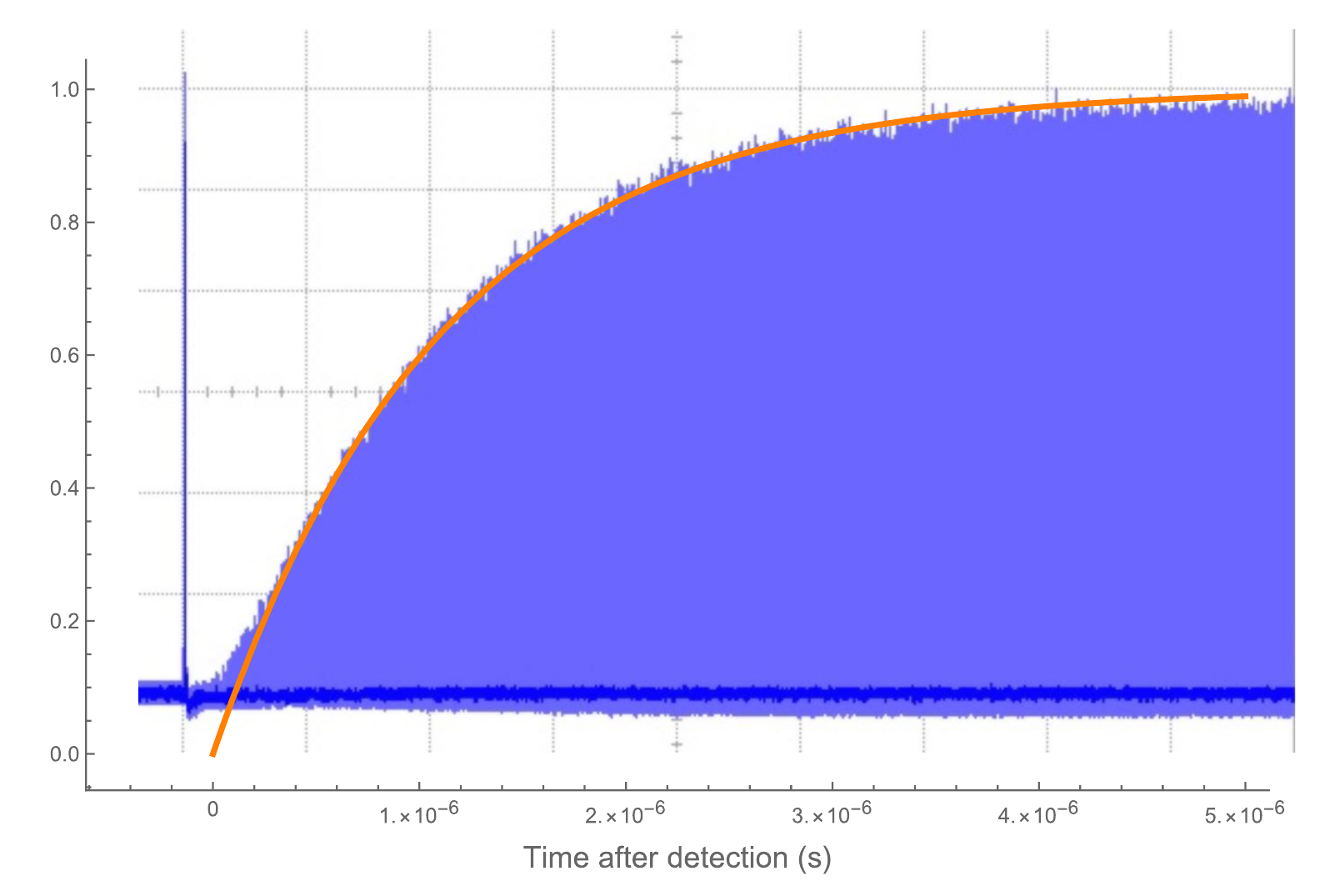}
\caption{ }\label{fig:rechargeExperiment}
\end{subfigure}
\hfill
\begin{subfigure}[b]{0.49\textwidth}
\includegraphics[width=\textwidth]{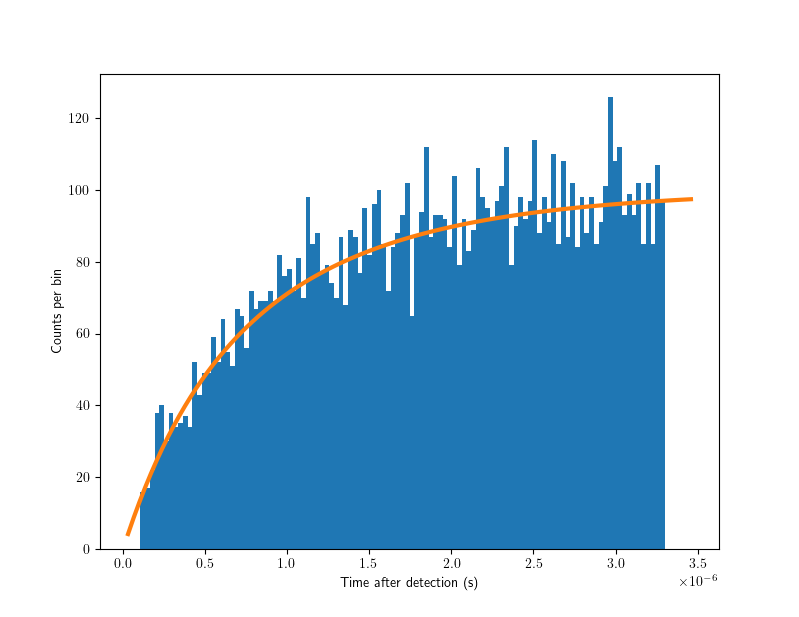}
\caption{ }\label{fig:rechargeSimulation}
\end{subfigure}
\caption{Experimental recharge curves and simulated analogue for passively quenched detectors. (A)~Experimental recharge curve from a representative diode collected with an oscilloscope in persistence mode, similar to \cite{DSouza2018}. A recharge curve following eq. \eqref{eq:defRechargeTime} with time constant $1.1\mu$s is overlaid, showing excellent agreement. (B)~Simulated data with the same time constant. A histogram of the first differences in timetags is presented. The deadtime ($\sim30$~ns) is visible near zero. The overlaid curve follows eq. \eqref{eq:rechargeDoubleExp}.}\label{fig:rechargeCurve}
\end{figure}

Figures~\ref{fig:rechargeCurve} shows experimentally measured recharge, compared with simulated recharge. The experimental data was obtained with a passive detector prototype, with an oscilloscope in persistence mode \cite{DSouza2018}. The simulated data was generated using the same example pass, now with the recharge time for one detector changed to match the experimental data ($1.1\,\mu$s compared to $700$ ns). The first difference of the recorded timetags is taken, and a histogram is plotted in Fig.~\ref{fig:rechargeSimulation}. For both experimental and simulated recharge curves, the relevant recharge equation is overlaid. Excellent agreement is observed between the recharge curves and the relevant data. 

\subsection{Including Eve}\label{sec:ResultsEve}

\begin{figure}[!htbp]
\centering
\includegraphics[width=\columnwidth]{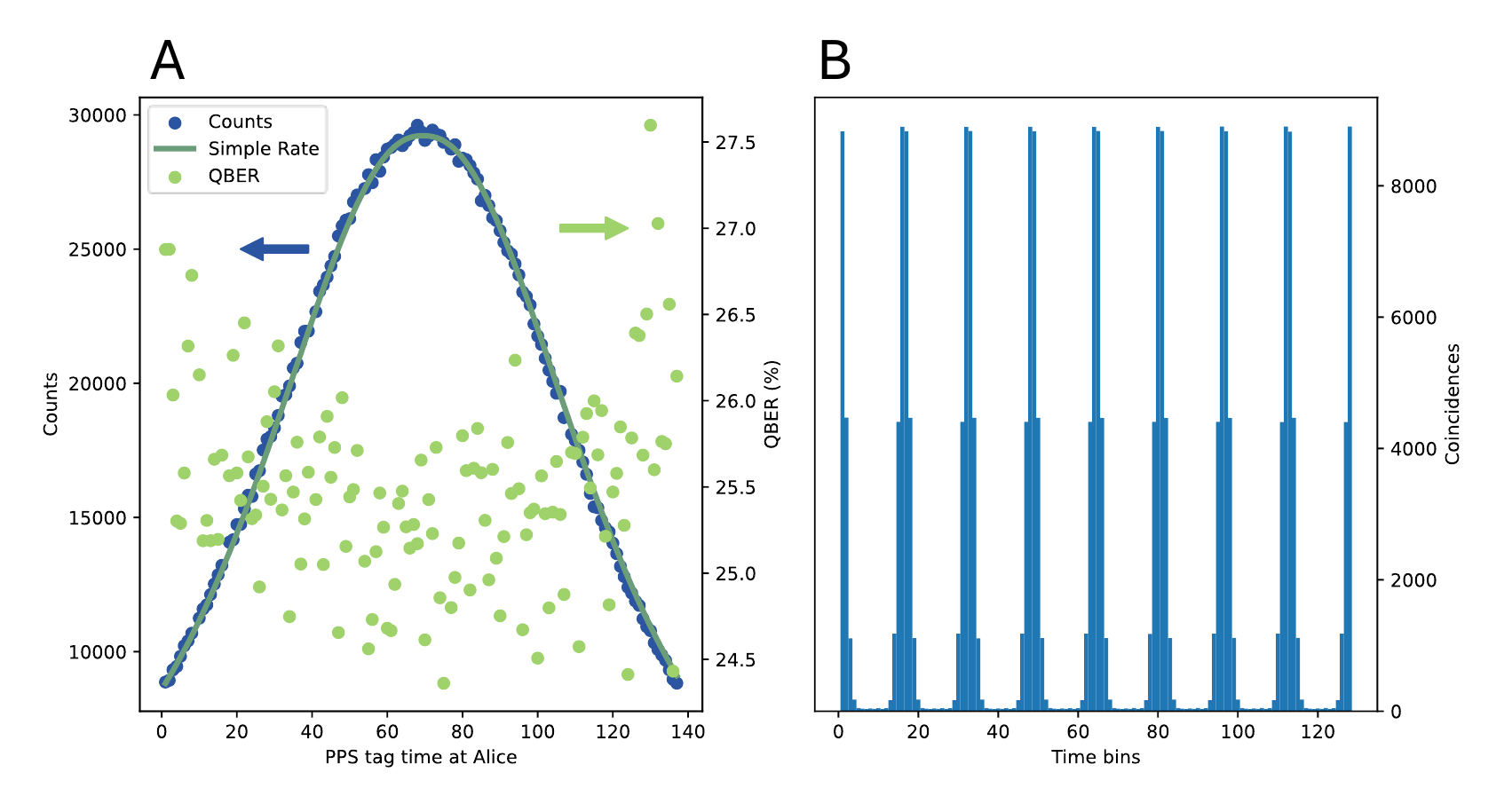}
\caption{Example of Eve's intercept-resend attack on observed statistics. (A) QBER is averaging 25.5\% QBER, which is beyond the threshold to distill a secure key. Even worse QBER can be observed at the edges of the pass, where noisy effects get more pronounced. (B) The coincidence histogram at frame 63 of figure (A). It does not show any noticeable differences compared to the un-eavesdropped histogram in Fig.~\ref{fig:qberSinglesHist}~(B), highlighting the power of Eve's intercept-resend attack.}\label{fig:EveQBERHist}
\end{figure}

Figure \ref{fig:EveQBERHist}~(A) shows the QBER obtained for the measured signals at Bob, where Eve intercepted 100\% of the emitted signals and is located at 3\% of the distance between Alice (0) and Bob (1). Both of these parameters can be tuned in the simulator. As expected for a full intercept-resend attack, Bob measures $\sim 25\%$ QBER. It is expected that the edges of the pass have worse performance even in the trusted scenario, which explains the higher QBER at the beginning and end of the pass. No key can be distilled, as the QBER is beyond the BB84 threshold of 11\%. 

\subsection{Application to QEYSSat: Optimization}\label{sec:optimAndAutomate}

\begin{figure}[!htbp]
      \centering
      \includegraphics[width=0.6\textwidth]{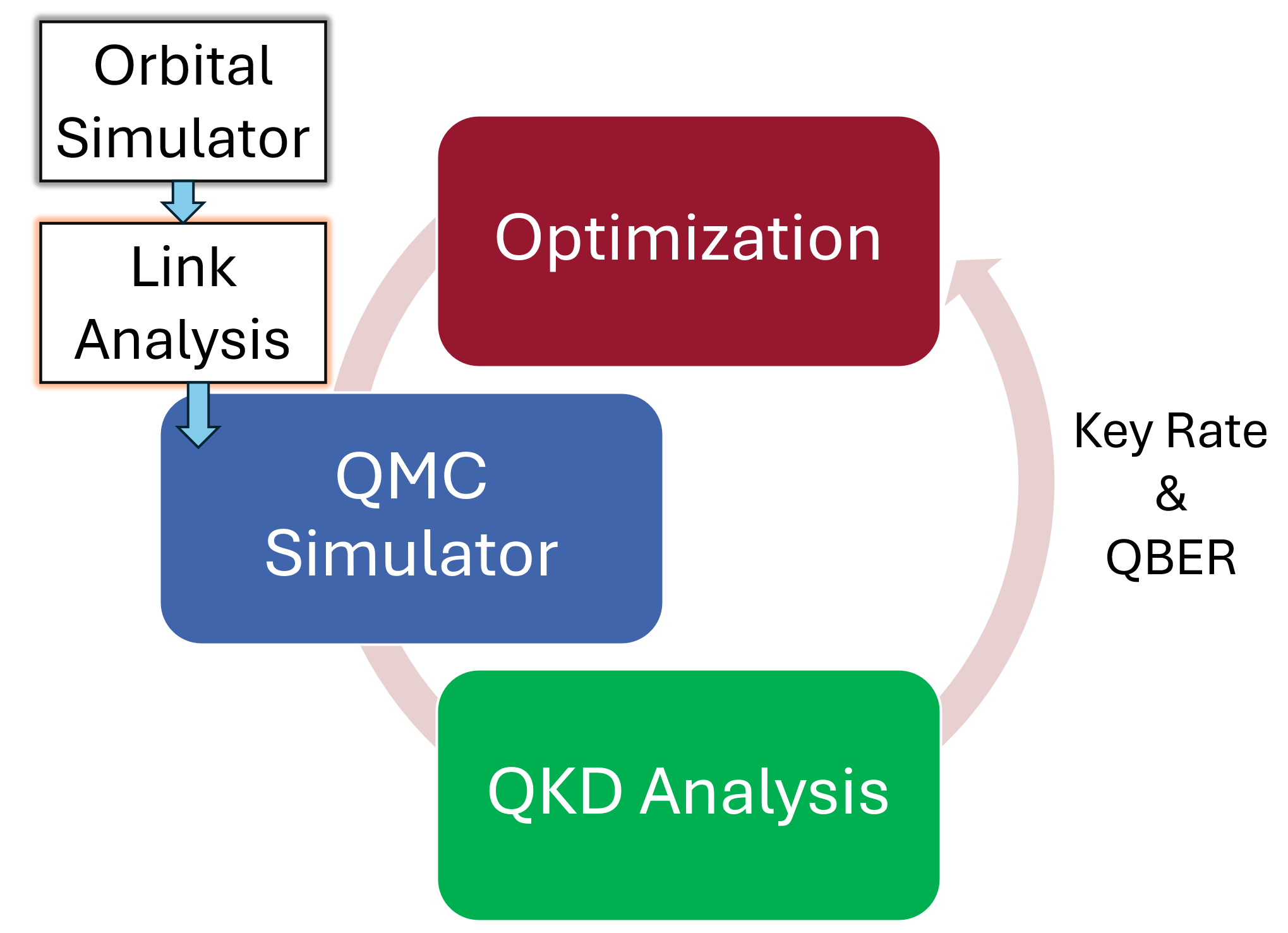}%
      \caption
        {%
          Automation workflow. All software components of the workflow are highlighted and the loop itself is visualized. Different satellite passes are generated in the orbital simulator (Ansys STK) and processed through our Link Analysis software. This data is fed into the QEYSSat Monte Carlo Simulator. A simulation is analyzed through the QKD Analysis software. Specific metrics (QBER and secret key length) are then extracted and used to update the objective function. The optimization routines return a new set of parameters which define the next scenario to simulate, and so on. As orbital pass information is an additional parameter, different ground stations can be compared. 
          \label{fig:optimizationLoop}%
        }%
\end{figure}

The QEYSSat mission presents a good use case for the QMC Simulator. For the mission to be successful, all of the parameters on and off the satellite must be optimized prior to launch. Some parameters will be accessible and adjustable after launch, and thus will provide valuable feedback and validation for our models. As a first step towards deploying the QMC Simulator for experimental validation, our group sought to optimize the values of two key parameters for the mission: (1) mean photon rate at the source, and (2) excess bias voltage applied to the detector. Two objective functions were chosen for this analysis: Final Key Rate and QBER. Technically, the optimization operation is an $\operatorname{argmax}$ or $\operatorname{argmin}$ evaluation for the ideal parameter values as a function of Final Key Length or QBER. 

Recalling that the QMCS does not use an overall analytic expression for the phenomena it models, this makes it a \textit{black box} objective function in the context of optimization. The choice of black-box algorithm was made carefully, comparing several options including a grid search, a linear Model-based Trust Region (MBTR) algorithm called COBYLA, a quadratic MBTR called LINCOA, and finally a Mesh-adaptive Direct Search (MADS) algorithm. The COBYLA algorithm demonstrated the best performance given the parameters and objective functions we used. It is suspected that as the number of parameters grows, the quadratic MBTR will outperform the linear MBTR algorithm. 

The QMCS must be run many times in the course of an optimization since numerous points in the parameter space must be probed to obtain objective values as the algorithm traces a path towards the optimum. We have developed a proprietary optimization software allowing us to automatically run an optimization algorithm over a user-defined parameter space, instead of manually entering all parameter values for every iteration the optimization algorithm may require. This optimization software currently incorporates 49 variable parameters, and the four optimization algorithms listed earlier.  

A given scenario of interest is structured in an array of \texttt{MATLAB} objects, such that each parameter which an end user may want to change is programmatically accessible. This allows for on-the-fly definition of scenarios, which can then be externally saved and simulated. This feature allows for automation of various scenarios, as well as enabling optimization of any given set of parameters.

Several optimization algorithms are implemented in an independent module \cite{Maierean2024} from the simulator. These algorithms are linked to the proprietary analysis software, which can produce the metrics required to evaluate the given objective functions, such as the QBER and the secret key length. Figure \ref{fig:optimizationLoop} illustrates how each software interacts with each other given the optimization task.

\section{Discussion}
\label{sec:discussion}

The QEYSSat Monte Carlo Simulator was initially discussed in Refs \cite{DSouza2018,Maierean2024}. The former introduced the major algorithms implemented, and the later enhanced the discussion around the optimization of given parameters for the mission. 

Development of the simulator was primarily used to cross-validate the QKD Analysis software which is being enhanced for the mission. It allowed generation of realistic data files using the same format as the payload will generate. Both the start and stop time within a pass are user parameters, allowing a user to simulate only a region of interest, such as the middle of the pass where the counts will be higher, and where the orbital speeds are different. 

Development and testing were performed on local computers. Full-length satellite passes were simulated using Digital Research Alliance of Canada clusters. These systems will be exploited for optimization purposes both prior and during the mission. While it is possible for a commercial computer to simulate a full pass, it does not perform ideally. The performance of the simulator is highly correlated with hardware, as we observed a large variance in the run time for the same test simulations depending on the specific machine used. Additional performance improvements are still being found this late into development. 

To generate the data required for the afterpulsing analysis, the simulator had to be run in an unusual configuration, that is, not simulating an orbital pass. Specifically, we had to let the detectors run without any additional source or background photons to accurately model the effect of dark counts and afterpulsing. This required letting the simulator run for an order of magnitude longer than the design calls for, and would be expected in practical use. This edge case demonstrated an interpolation function was the source of an unexpectedly long execution time. Vectorizing this function and the performance of the nominal scenario also improved, from an acceptable 1h40 minutes down to a record 43 minutes for the same baseline scenario presented in Fig. \ref{fig:qberSinglesHist}.

Random number generation has been identified as the limiting factor for the performance of the simulator. These numbers are generated on the fly to simplify the structure of the function calls. Care was taken to optimize algorithms where possible to exploit \texttt{MATLAB}'s features. Similar to the random sequence of states being pre-generated before a true pass, pre-generating the random numbers used in the QMC Simulator may improve performance. 

A priority use case for the simulator will be to determine what operational parameters to use prior to a data collection pass. Thus, the automation and optimization features as highlighted in Sec.~\ref{sec:optimAndAutomate} of the simulator will be critical to achieve this. 

The main phase of QEYSSat experiments include the WCP source demonstrating decoy-state BB84, and an EPS demonstrating BBM92. Further planned experiments include a single photon source, e.g., a nanowire quantum dot \cite{Reimer2012,Pennacchietti2024}. Adding different sources requires knowledge of the photon number probability distribution of the source. The new source class would therefore generate photon numbers according to this distribution, convert them to timetags, and pass them to the channel module. This feature highlights the adaptability of our formalism. 

As the simulator ingests link profiles, different atmospheric prediction models can be compared using the same experimental setup. Furthermore, experimentally measured atmospheric conditions can be used to generate link profiles after the fact, allowing for the simulation of observed, rather than predicted, conditions. Generating data for different QEYSSat ground stations and including them in the automation process fulfills one of the key tasks for the simulator. On any given pass, QEYSSat may be able to link with a number of possible ground stations. Given weather conditions at each ground station, different performances are expected. The QMCS will be used to evaluate which ground station should be used based on expected performance.

Using a look up table, the QMCS is able to control the efficiency and dark count rates, as a function of the excess bias voltage and temperatures of the detectors. These variables will be key parameters to optimize during the mission. This level of control, based on the level of irradiation expected \cite{Anisimova2021,DSouza2021}, allows the simulation of expected conditions both at the beginning and at the end of the mission.

\section{Outlook}\label{sec:outlook}
We presented a novel Monte Carlo based photon emission, transmission, and detection simulator for the use case of satellite-based quantum key distribution. This probabilistic simulation allows for the inclusion of memory-based effects such as afterpulsing, which are hard to include analytically. The effect of variations in specific parameters over the final secure key length can be determined, and used to determine operational parameters for experimental satellite passes. 

While the baseline weak coherent pulse source has been highlighted here, further work will include the full simulation of the entangled photon emitter, other sources such as single-photon emitters, and improving the workflow such as removing the remaining man-in-the-middle link in the automation process. Another important future enhancement of our simulator is to include a detailed  models - or the actual measured link performance - of transmission profiles and fast scintillation. In addition, after launch, the simulator will be used to validate experimental data with predicted simulations, validate related link models, and be continuously used to determine operational parameters as the detectors degrade due to irradiation in space \cite{Anisimova2017,DSouza2021}. By comparing simulations and experiments we plan to validate and improve this digital twin solution. Development on the simulator will thus continue past the satellite launch, as it will be updated to implement and reflect the experimental results. We anticipate that our high-fidelity digital twin of the QEYSSat quantum key distribution system will help optimize the scientific results from QEYSSat, and provide a baseline for sophisticated models for future quantum communication satellite missions.

\begin{acknowledgments}
The authors thank the Canadian Space Agency (CSA), the Natural Sciences and Engineering Research Council (NSERC) of Canada, the Ontario Research Fund, the National Research Council (NRC) High Throughput and Secure Networks (HTSN) program, and the Canada Excellence Research Chair (CERC) program for funding. HPNM acknowledges support from the NSERC CGRS Doctoral and Ontario Graduate Scholarship programs. This research was enabled in part by support provided by Compute Ontario and the Digital Research Alliance of Canada. 

The authors thank the numerous contributors to the simulator code over many years of development: Simon Friesen, Sebastian Slaman, Sebastian Ruan, Alissa Van Gaalen, Dean Fountas, Khasir Hean, Tudor Mantaila, Eshaan Raval, and Evan Blaylock. We also thank the following individuals for discussions and support: J.-P. Bourgoin, Paul Godin, Alexander Koujelev, Norbert L\"utkenhaus, Brian Moffat, Kimia Mohammadi,  Sungeun Oh, Lydia Philpott.
\end{acknowledgments}

\bibliography{sample}

\end{document}